\documentstyle[12pt]{article}
\def\cmm2{{\,\rm cm^{-2}}}
\def\cm2{{\,{\rm cm}^2}}
\def\cmm3{{\,{\rm cm}^{-3}}}
\def\gcmm3{{\,{\rm g\,cm^{-3}}}}

\def\fun#1#2{\lower3.6pt\vbox{\baselineskip0pt\lineskip.9pt
  \ialign{$\mathsurround=0pt#1\hfil##\hfil$\crcr#2\crcr\sim\crcr}}}

\begin{document}
\baselineskip=18pt
\begin{center}
\bigskip

\rightline{CWRU-P13-96}

\vspace{0.5in}
{\Large \bf Old Galaxies at High Redshift and the Cosmological
Constant}
\vspace{0.35in}

Lawrence M. Krauss\\
\vspace{.1in}
\baselineskip=16pt
{\it Departments of Physics and Astronomy \\
Case Western Reserve University\\
10900 Euclid Ave. \\
Cleveland, OH~~44106-7079}\\

\end{center}

\vspace{.3in}
\baselineskip=14pt

{\bf In a recent striking discovery, Dunlop {\bf \it et al} \cite{dunlop}
observed a galaxy at redshift z=1.55 with an estimated age of 3.5 Gyr.  This is
incompatible with age estimates for a flat matter dominated universe unless the
Hubble constant is less than $ 45 kms^{-1}Mpc^{-1}$.  While both an open
universe, and a universe with a cosmological constant alleviate this problem, I
argue here that this result favors a non-zero cosmological constant, especially
when considered in light of other cosmological constraints.  In the first place,
for the favored range of matter densities, this
constraint is more stringent than the globular cluster age constraint, which
already favors a non-zero cosmological constant.  Moreover,  the age-redshift
relation for redshifts of order unity implies that the ratio between the age
associated with redshift 1.55 and the present age is also generally larger for a
cosmological constant dominated universe than for an open universe. In addition,
structure formation is generally suppressed in low density cosmologies, arguing
against early galaxy formation. The additional constraints imposed by the new
observation on the parameter space of
$h$ vs
$\Omega_{matter}$ (where 
$H= 100 h kms^{-1}Mpc^{-1}$) 
are 
derived for both cosmologies. For a cosmological constant dominated universe
this constraint is consistent with the range allowed by other
cosmological constraints, which also favor a non-zero value \cite{mk}.
}  

\baselineskip=18pt

\vskip 0.2in
Whenever Big Bang cosmology has been challenged by an age problem, a cosmological
constant has been invoked
as a possible remedy(i.e. \cite{mk}).  The reason is simple.  As
long as the universe is decelerating, the present Hubble expansion rate sets an
upper limit on the age of the universe as follows.  If there were no
deceleration as the universe evolved, the distance galaxies would have travelled
since $t=0$ would be $d=vt$.  Since the Hubble constant $H=v/d$, then
$H^{-1}=t$. If the universe has been decelerating, then distant galaxies would
have achieved their present distances in a shorter time.  Thus, for any matter
(or radiation) dominated cosmology the age of the universe
$\tau < H^{-1}$.  If independent estimates of the ages of galaxies are
larger than this value, there is an apparent paradox. However the addition of a
cosmological constant allows a period of cosmic acceleration rather than
deceleration, and hence allows the obviation of this bound, and the paradox.
Recently it has been recognized that a number of other cosmological observables,
including the baryon density of the universe and the shape of the power
spectrum of galaxy-galaxy correlations, also argue in favor of a cosmological
constant, at least if one is to preserve a flat universe, renewing interest in
the possibility that the cosmological constantis non-zero, in spite of
the theoretical microphysical problems associated with this idea (i.e. see
\cite{mk, kofman}). 

These latter cosmological constraints do not distinguish between an open
universe, and a flat universe dominated by a cosmological constant, however. 
Indeed, the age problem, most recently quantified by the discrepancy between
the inferred ages of globular clusters \cite{chab} and the Hubble age, has
provided perhaps the strongest motivation for considering one cosmology over the
other (although COBE-normalized density fluctuations also are significant in
this regard, as I shall describe later).   The recent discovery of a 3.5-Gyr-old
galaxy at a redshift of 1.55 is therefore particularly interesting in this
regard. 

Of interest in this case is the effect of a cosmological constant not on the
present age, but rather the age of the universe at redshifts of order unity. 
It is clear that a cosmological constant which is significant today will only
affect the evolution of the universe for low redshifts.  However, it is well
known that a cosmological constant alters the distance-redshift
relation for redshifts of order unity----associating longer distances with a
given redshift---enough to dramatically affect such things as the optical depth
for gravitational lensing of distant quasars
\cite{tf,kw,koch}. For the same reason, one might expect that for low z, the
age of the universe may be lengthened significantly compared to a flat matter
dominated universe or an open universe. 

The age of a flat matter dominated universe with Hubble constant $H$ is
given by
$ \tau =(2/3) H^{-1}$.  Since $H$ goes as $R^{-3/2}$ during the matter dominated
era, the age-redshift relation is trivially given as $ \tau \approx
(1+z)^{-3/2} $ The z-dependent age for an matter dominated open universe and for
a cosmological constant- dominated flat universe are somewhat more complicated,
but nevertheless can be straightforwardly derived and expressed in terms of the
present Hubble constant
$H_0$ as follows:

\begin{equation} 
Open: ~\tau (z) = H_0^{-1} \int_{0}^{(1+z)^-1} (1+ \Omega_{0,matter}
+\Omega_{0,matter} x^{-1})^{-1/2}
\ dx 
\end{equation}

\begin{equation}
\Lambda: ~\tau (z)= (2/3) H_0^{-1} \Omega_{0,\Lambda}^{-1/2} ln[(( {\rho(z)
\over
\rho_0} + {\Omega_{0,\Lambda} \over \Omega_{0,matter}})^{1/2} +
({\Omega_{0,\Lambda} \over
\Omega_{0,matter}})^{1/2}) ({\rho (z) \over \rho_0})^{-1/2}]  
\end{equation}
where $\Omega_{0,\Lambda},\Omega_{0,matter}$ are the
fractional contribution of the cosmological constant today and the matter
density today, compared to the closure density respectively, and $\rho_0$ is the
energy density of matter at the present time.

Many independent estimators, including virial estimates of cluster masses, X-Ray
estimates of the total mass in clusters of galaxies, and large scale velocity
field measures, suggest that $\Omega_{0,matter} \ge 0.2$.  The equations above
then imply that $\tau_0 < 8.27/h) Gyr$ for an open universe, and $\tau_0 <
(10.46/h) Gyr$ for a cosmological constant dominated flat universe, vs
$\tau_0 < (6.52/h) Gyr$ for a flat matter dominated universe.  A comparison
of these ages with a 95$\% $ confidence
lower bound on the age of the oldest globular clusters in our galaxy, $\tau_0
\ge 12.1 Gyr$, from a recent 
comprehensive analysis of theoretical and observational uncertainties in
globular cluster age estimates \cite{chab} provides a quantitative measure of
the current cosmological ``age problem".

To what extent does the recent Dunlop {\it et al} observation impact on this
situation?  Setting $z=1.55$ in the above equations, one finds 
$\tau < 2.67/h ~Gyr$ and
$ \tau < 3.53/h ~Gyr $ for an open, and cosmological constant dominated flat
universe respectively.   
These relations clearly indicate that values
of $h$ which satisfy the current cosmological age problem can also result in
cosmological ages at $z=1.55$ greater than 3.5 Gyr for both cosmological
models.  However they also provide greater insight into the relative viability
of both models, especially when other cosmological constraints are taken into
account, as is the full allowed range of $\Omega_{0,matter}$.   In
the first place, note, that not only is the absolute age of the
universe at this redshift larger in a cosmological constant dominated model
than in an open universe model for this value of the matter density, but that
the ratio of ages is larger at $z=1.55$ than for $z=0$.  This situation
persists as long as $\Omega_{0,matter} \le 0.5$.  Moreover, while for
$\Omega_{0,matter} =0.2$ the $z=1.55$ observation provides a weaker constraint
on $h$ than the current globular cluster age estimate does, this situation
quickly changes for larger values of $\Omega_{0,matter}$.

To fully appreciate the significance of these issues, it is useful to plot the
constraint on the full $h, \Omega_{0,matter}$ parameter space implied by 
$\tau (z=1.55) \ge 3.5 Gyr$, for both open and flat $\Lambda$
dominated cosmologies, along with constraints from other cosmological
measurements, following the approach of an earlier analysis for the flat
$\Lambda$ model \cite{mk}.  Such constraints are presented in figures 1 a and
b.  The line representing the Dunlop {\it et al} limit is explicitly labeled,
and the viable region is constrained to be below this line.  The other
constraints arise as follows: region (c) represents the globular cluster age
constraint
$ 12 Gyr < \tau_0 < 18 Gyr$ \cite{chab}; region (a) arises from considerations
of the baryon content of the universe determined by reconciling estimates from
Big Bang Nucleosynthesis \cite{kk3,cst}, which suggest that $.009 \le \Omega_B
h^2 \le .022$,  with determinations based on X-Ray measurements of rich clusters
of galaxies
\cite{white,dwhite}, which suggest $ \Omega_B / \Omega_{cluster} > 0.05-.08
h^{-3/2}$.  Combining the two together, and assuming $\Omega_{cluster} \approx
\Omega_{matter}$, one derives the constraint $ 0.11 < \Omega_{matter} h^{1/2} <
0.44 $; region (b) corresponds to the constraint coming from observations of
the shape of power spectrum of galaxy clustering, which yield
\cite{kofman,peak,costa}  
$0.2 < \Omega_{matter} h < 0.3$ ; the dashed  curves in figure 1(a) give limits
on the allowed parameter space derived from matching the COBE normalized
fluctuations in the CBR to inferred density fluctuations on galaxy scales in a
flat $\Lambda$ dominated universe in Cold Dark Matter models \cite{bunnwhite}. 
Finally, in figure 1(a) a bound (d) is shown which corresponds to
the assumption that
$\Omega_{matter} \ge 0.3$ for a flat $\Lambda$ dominated universe.  This bound
arises in part from considerations of gravitational lensing probabilities
\cite{tf, kw,koch}.  It is also consistent with the fact that dynamical
estimates of the clustered mass on large scales actually generally favor
$\Omega_{matter} > 0.3$, rather than the more conservative estimate of 0.2
mentioned earlier.

While the latter bound is not included explicitly in figure 1(b), corresponding
to an open universe, it is clear that the age constraint coming from globular
clusters, combined with the power spectrum constraint, together imply a joint
allowed region in which $\Omega_{matter} > 0.3$ in any case.  Note that it is
precisely in this allowed region that the new Dunlop {\it et al} provides a
tighter parameter space constraint than the globular cluster age
limit.  Thus, for all effective purposes, in an open universe this new
constraint is actually stronger than the previous well known globular age
constraint.  Note also, that a noticeable fraction of the previously allowed
range of parameter space in the case of an open universe is now excluded.  By
comparison, the Dunlop {\it et al} constraint is only marginally stronger
than the globular cluster age constraint for a flat $\Lambda$ dominated
universe, and essentially all of the parameter space which was previously
allowed is still allowed.

Also note that the constraint from COBE which appears in figure 1(a) is not
included on figure 1 (b).  This is because the COBE normalized density
fluctuations generally are more difficult to fit to observed density fluctuations
on galactic scales in low density open universe models \cite{kofman,costa}.  This
is because the growth of density fluctuations since recombination is suppressed
in low density models compared to $\Lambda$ dominated models because the former
become curvature dominated at earlier redshifts than the latter become
$\Lambda$ dominated \cite{kofman,bunnwhite}. 
Moreover, and perhaps more important in the context of this discussion, because
the growth of fluctuations is suppressed, galaxy formation will tend to occur
later in low a low density universe, at least one with an initial
relativity flat spectrum of density fluctuations.  This in itself tends to argue
against the formation of galaxies as old as 3.5 Gyr at a redshift of 1.55 in
such models. 

In conclusion, the new Dunlop {\it et al ~} observation provides a more
severe constraint on open universe models than it does on cosmological
constant dominated flat models. It is completely
consistent with previous cosmological constraints on $\Lambda$ dominated
models, while it noticeably reduces the allowed $h -\Omega_{matter}$ parameter
space for open models.  In particular, for the range of parameter space which
was favored by other cosmological constraints, the $z=1.55$ age limit is
more powerful than existing globular cluster age constraints.  In other
words, it provides an even more severe ``age problem" for such cosmologies. 
Arguments based on structure formation in light of the COBE results reinforce
this favoring of $\Lambda$ dominated models in the context of this new result
because they generally allow earlier galaxy formation in these models. 

It is very important to note, however that the constraints here, being
cosmological, may best be considered as suggestive.  It is
possible that any one of them could be subject to large, as of yet
unanticipated systematic shifts.  However, at present the data seems to
point in at least one consistent direction. And on the face of it, the
observation of a $3.5 Gyr$ old galaxy at a redshift of 1.55 seems to provide
additional evidence favoring a cosmological constant dominated universe.  In
particular, if $h> 0.65$ it will be very difficult for any other cosmology to
satisfy both age constraints.  Also, if other old galaxies are discovered
at similar, or higher redshifts, the only model with a non-zero range of allowed
parameter space for any value of the hubble constant may be that in
which the cosmological constant is non-zero.  For this reason, independent
probes of the cosmological constant, including direct measures of
$q_0$ from Type 1a supernovae \cite{perl}, and full sky measurements of CBR
anisotropies on small angular scales take on even greater interest.

\newpage

\noindent{\bf Figure Captions}

Figure 1 (a)  Constraints on the parameter space of $h (=H_0/100 km
s^{-1}Mpc^{-1}$ vs
$\Omega_{matter}$ for a flat $\Lambda$ dominated universe coming from diverse
cosmological observations.  The bold line represents the limit of the region
excluded by the recent observation of a 3.5 Gyr old galaxy at a redshift of
1.55. Region (a) represents a bound coming from a comparsion of Big Bang
Nucleosynthesis predictions with X-ray observations of rich clusters. Region
(b) represents the constraint imposed by the shape of the power spectrum of
matter density perturbations on galaxy scales. Region (c) comes from
constraints on the age of globular clusters. The bound (d) arise both from
observations of gravitational lensing, and dynamical estimates of the clustered
mass in the universe.  The dashed lines represent the limits of the regions
allowed by a comparison of COBE CBR fluctuations with matter density
fluctuations for CDM models.

Figure 1(b)  Same as (a), except for the case of an open universe.  Regions
(a-c) are based on the same constraints as displayed in figure 1(a).

\newpage

\begin{figure}[htb]
\vglue 6.0in
\includegraphics{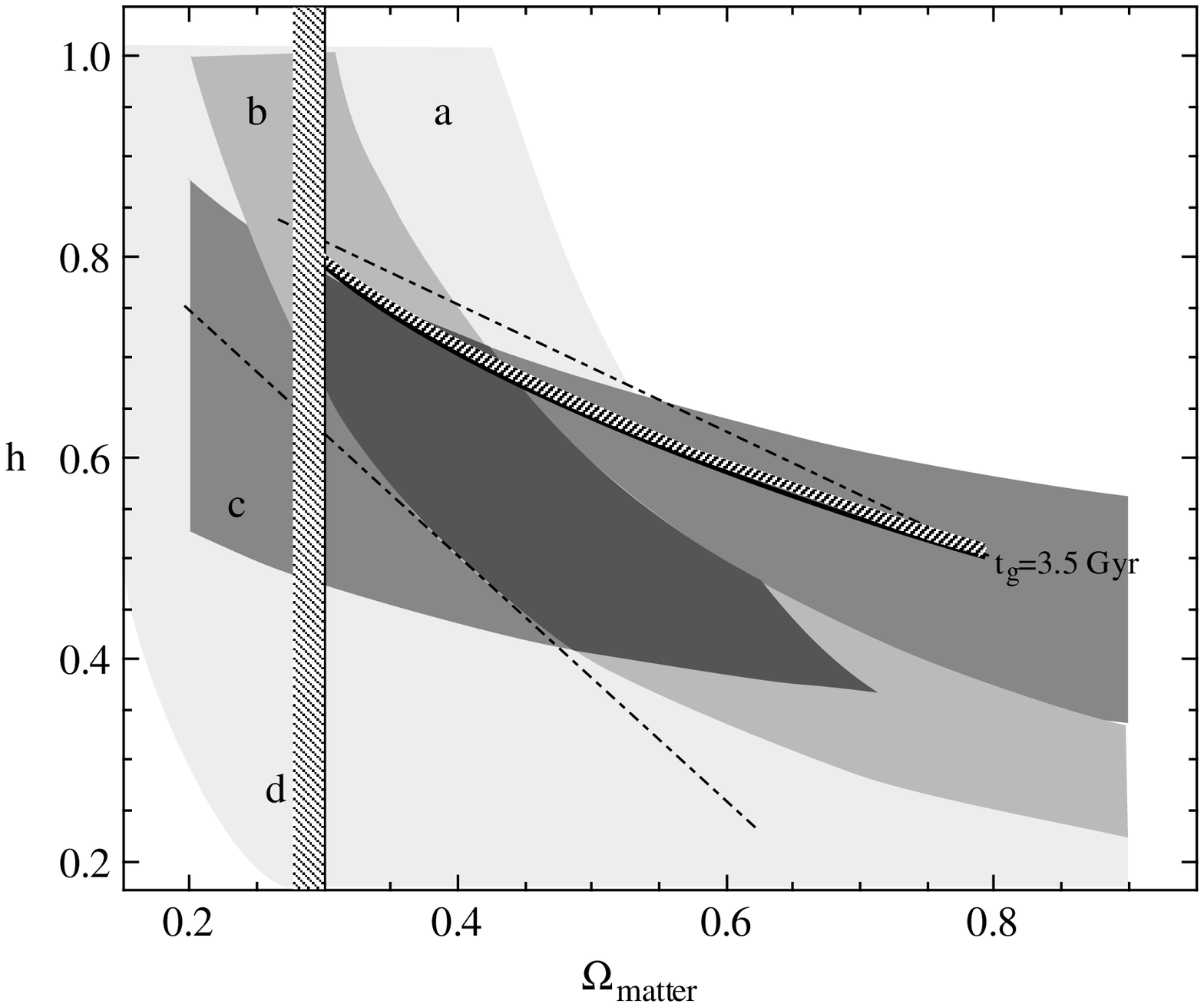}
Figure 1(a) Constraints on $\Omega_{matter}$ and $h$ for a 
cosmological constant
dominated universe
\label{fig:cosmo}
\end{figure}

\newpage
\begin{figure}[htb]
\vglue 6.0in
\includegraphics{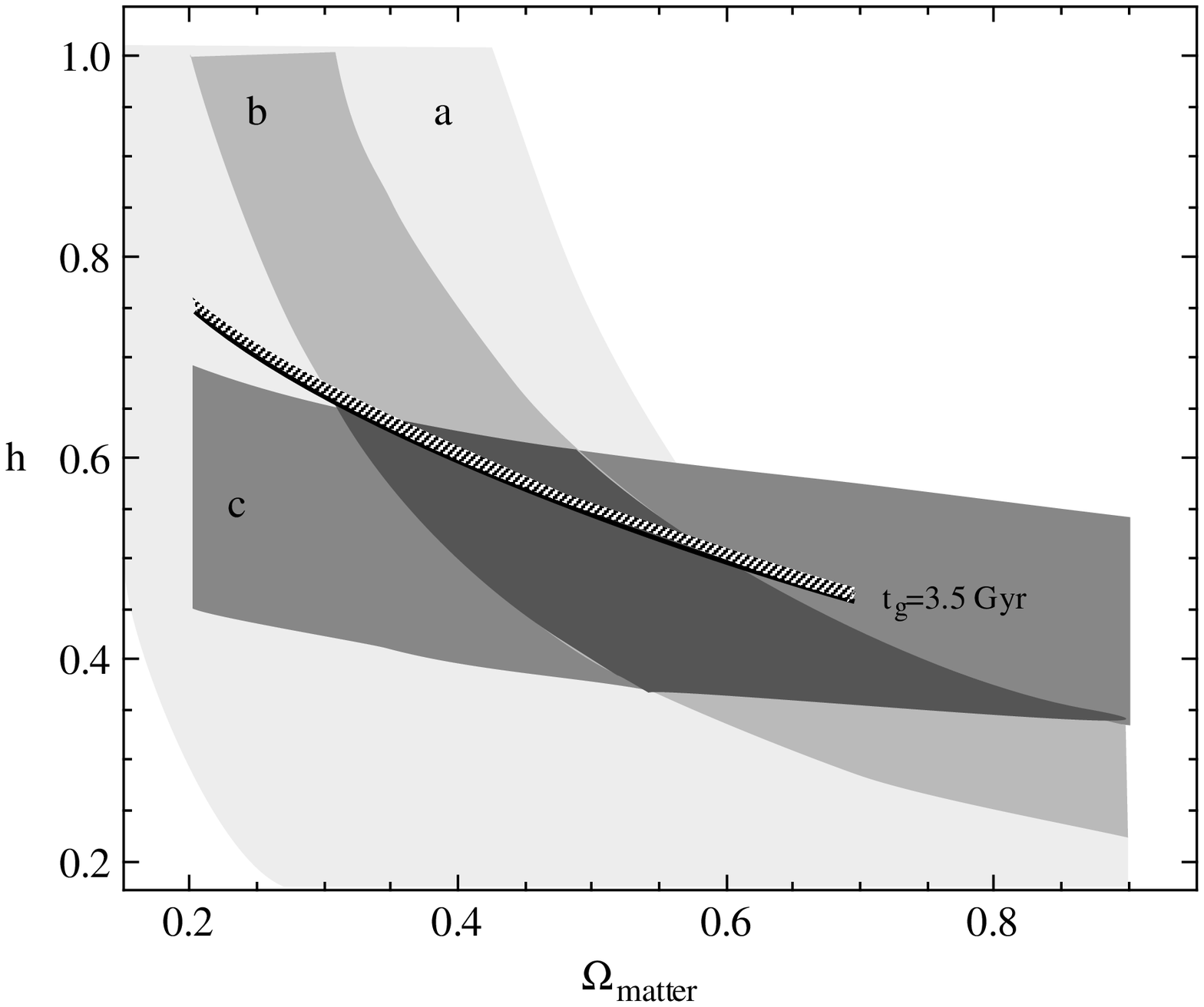}
Figure 1(b) Constraints on $\Omega_{matter}$ and $h$ for an 
open universe
\label{fig:open}
\end{figure}


\begin{thebibliography}  {gcages}
\baselineskip=15pt
\bibitem{dunlop} J. Dunlop {\it et al}, {\it Nature} {\bf 381} 581-584 (1996)

\bibitem{mk} L. M. Krauss and M. S. Turner, {\it J. Gen. Rel. Grav.} 
{\bf 27} 1137-1144 (1995) 

\bibitem{kofman} L. Kofman, N. Y. Gnedin and N.A. Bahcall, {\it Ap. J.}
{\bf413}, 1 (1993) 

 \bibitem{chab} B. Chaboyer, P. Demarque, P. J. Kernan, L. M. Krauss, 
{\it Science}, {\bf 271}, 957 (1996) 

\bibitem{tf}  M. Fukujita and E. L. Turner, {\it MNRAS} {\bf 253}, 99
(1991)

\bibitem{kw}  L.M. Krauss and M. White, {\it Ap. J. } {\bf 394} 385
(1992)

\bibitem{koch} C. S. Kochanek {\it Ap.J.} {\bf 348}, 1 (1992)


 \bibitem{kk3} L. M. Krauss and P.J. Kernan, {\it Phys. Lett.} 
{\bf B347}, 347 (1995) 

  \bibitem{cst} C. Copi, D.N. Schramm, and M. Turner, {\it Science}
{\bf 267}, 192 (1995)

\bibitem{white} S.D.M.~White {\it et al}, {\it Nature} {\bf 366},
429 (1993)

  \bibitem{dwhite} D. A. White and A. C. Fabian, {\it M.N.R.A.S.}
{\bf 273}, 73 (1995)

\bibitem{peak} J.A.~Peacock and S.J.~Dodds, 
{\it Mon. Not. R. astr. Soc.} {\bf 267}, 1020 (1994)

\bibitem{costa} L. N. da Costa {\em et al}, {\it Ap. J. Lett.} {\bf
 437} 97 (1994)

\bibitem{bunnwhite}  M. White and E. F. Bunn {\it Ap. J.} { \bf 450}
477 (1995)


\bibitem{perl} S. Perlmutter {\it et al}, LBL preprint (1995)


\end{thebibliography}
\end{document}